\documentclass[11pt, journal,transmag]{IEEEtran}
\usepackage{cite}
\usepackage[dvips]{graphicx}
\usepackage{amsmath}
\usepackage{amssymb}
\usepackage{array}
\usepackage[]{algorithm2e}
\usepackage{subcaption}
\usepackage{longtable}

\makeatletter
\def\BState{\State\hskip-\ALG@thistlm}
\makeatother
\usepackage{xargs}
\usepackage{booktabs}
\usepackage[pdftex,dvipsnames,table,xcdraw]{xcolor} 
\usepackage[colorinlistoftodos,prependcaption,textsize=tiny]{todonotes}
\newcommandx{\unsure}[2][1=]{\todo[linecolor=red,backgroundcolor=red!25,bordercolor=red,#1]{#2}}
\newcommandx{\change}[2][1=]{\todo[linecolor=blue,backgroundcolor=blue!25,bordercolor=blue,#1]{#2}}
\newcommandx{\info}[2][1=]{\todo[linecolor=OliveGreen,backgroundcolor=OliveGreen!25,bordercolor=OliveGreen,#1]{#2}}
\newcommandx{\improvement}[2][1=]{\todo[inline,linecolor=Plum,backgroundcolor=Plum!25,bordercolor=Plum,#1]{#2}}
\newcommandx{\thiswillnotshow}[2][1=]{\todo[disable,#1]{#2}}

\graphicspath{{./figures/}}
\usepackage[utf8]{inputenc}
\begin{document}

%

\title{Density-aware Dynamic Mobile Networks:\\Opportunities and Challenges}

\author{\IEEEauthorblockN{
Ertan Onur, 
Shahram Mollahasani,
Alperen Ero\u{g}lu and Nina Razi Moftakhar
}
\IEEEauthorblockA{}
\thanks{Corresponding author: E. Onur (email: eronur@metu.edu.tr).}}

%

\IEEEtitleabstractindextext{%
\begin{abstract}
We experience a major paradigm change in mobile networks. The infrastructure of cellular networks becomes mobile as it is densified by using mobile and nomadic small cells to increase coverage and capacity. Furthermore, the innovative approaches such as green operation through sleep scheduling, user-controlled small cells, and end-to-end slicing will make the network highly dynamic. Mobile cells, while  bringing many benefits, introduce many unconventional challenges that we present in this paper. We have to introduce novel techniques for adapting  network functions, communication protocols and their parameters to network density. Especially when cells on wheels or wings are considered, static and man-made configurations will waste valuable resources such as spectrum or energy if density is not considered as an optimization parameter. In this paper, we present the existing density estimators.  We analyze the impact of density on coverage, interference, mobility management, scalability, capacity, caching, routing protocols and energy consumption. We evaluate  nomadic cells in dynamic networks in a comprehensive way and illustrate the potential objectives we can achieve by adapting mobile networks to base station density. The main challenges we may face by employing dynamic networks and how we can tackle these problems are discussed in detail. 
\end{abstract}

}

\maketitle

\section{Introduction}

\IEEEPARstart{T}{he}  state of the art in mobile cellular network is the centrally-managed and relatively inflexible  architecture that was prosperous albeit not scalable any more. The present-day networks have already reached the spectrum limitations. We have  to densify cellular networks by spatial multiplexing to overcome capacity limitations. Increasing the number of base stations (BS) may  cause severe interference and redundant coverage resulting in energy wastage.  Centralized configuration or real-time centralized monitoring is not applicable due to the difficulties in acquiring global information about the network and computational complexity of the tasks. Management, coordination and optimization tasks usually require solving NP-hard problems. 

Nokia, in their white paper on ultra-dense networks\footnote{http://resources.nokia.com/asset/200295}, predicts that the number of base stations per squared kilometer will increase from $\sim$10 in 2014 to $\sim$100 per squared kilometer in densest places beyond 2020 and the inter-site distance will reduce to $\sim$100 from $\sim$400 meters. As the network enlarges, its management and control become a symptomatic issue. Operator intervention requirements have to be drastically reduced by employing self-organization. There is a research gap between the state of the art and the ambition of achieving a self-organized, adaptive and flexible networking architecture.  Moreover, we are at the verge of several key paradigm changes in mobile communications. 

\subsection{Paradigm Changes in Mobile Communications}
One of the significant paradigm shifts happens in the control domain of operators. In the past, network operators used to plan, dimension and install base stations. Before and after the launch of the base stations, optimization was plausible. Performance monitoring, failure mitigation and correction were carried out by the network operator within the lifetime of a base station. However, this scheme will change substantially in future mobile networks and operators will partially lose their control on cell deployment as we will explain in this paper.

\begin{figure*}[h]
\centering
	\includegraphics [width=\textwidth] {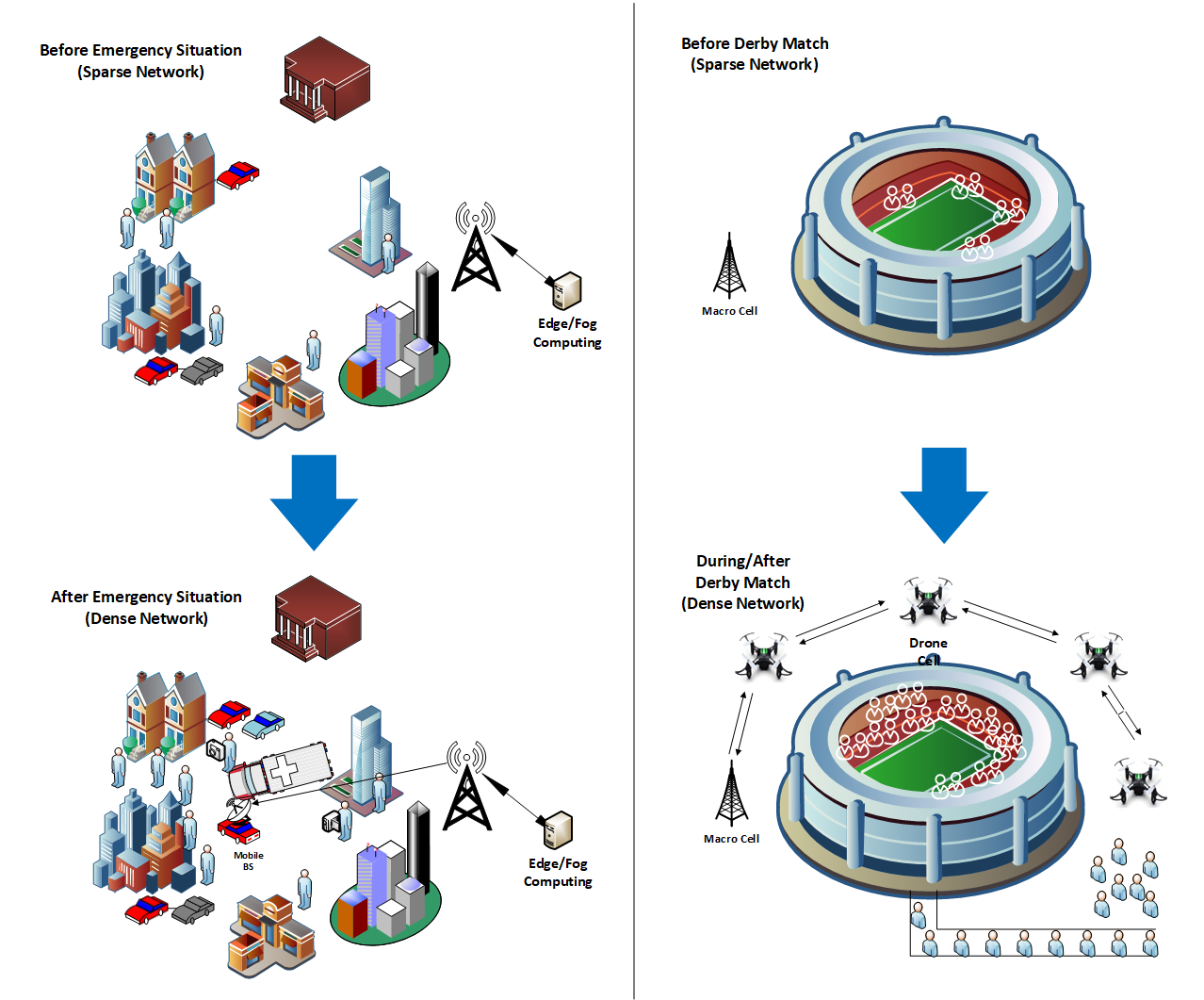}  
\caption{Two application scenarios of mobile base stations in  future networks are presented. We illustrate how cells on wings or wheels may change the infrastructure of mobile networks. Because of mobility and many other factors we present in this paper, the infrastructure of mobile networks start resembling ad hoc networks in terms of dynamics. As a consequence, the density of base stations  unpredictably change. }
\label{scenario1}
\end{figure*}

Another paradigm change is in the infrastructure of mobile networks. In the past, we used to assume that locations of user equipment (UE) were stochastic, and the network infrastructure was stationary. In the future, base stations may also be mobile yielding a random infrastructure; e.g., drones may provide service to blind spots~\cite{Yaliniz2016}.  We present some example scenarios where the density of UEs and also BSs may change in a dynamic fashion in Fig.~\ref{scenario1}.  As can be seen in these scenarios, the density of users may increase suddenly because of some emergency situation such as a car accident or a sports event. As we can see on the left hand side of Fig.~\ref{scenario1}, the area seems sparse initially. However, after the car accident, the density of users  increases dramatically. Therefore,  mobile or nomadic BSs are deployed in the area to maintain quality of service (QoS) in terms of coverage or capacity. In emergency cases, pre-deployment planning may not be possible. Communication services are of critical importance for public protection and disaster recovery. Man-made or natural disasters such as earthquakes may disrupt communication services that are currently provided by stationary infrastructures. Employing drone base stations can be a viable approach for establishing a communication infrastructure in affected areas and for providing wireless coverage in blind spots. Drone base stations can also be used for gathering data from rural fields where no communication infrastructure exists. For instance, drone cells may act as mobile sinks in applications of the Internet of Things and in massive machine type communication scenarios. 

As another scenario,  a derby football match can be given.  Some  flying BSs such as drones may provide coverage and enhance  QoS during the event as presented on the right hand side of Fig.~\ref{scenario1}. Before the event and after the event, the user density in the stadium will be low. However, it will be substantially larger during the match. Instead of incurring the cost of deploying stationary cells inside or nomadic cells around the stadium, cells on wings may be employed on the stadium to satisfy the QoS requirements of users by getting closer to UEs. Depending on the user density, additional base stations can be dynamically deployed that in turn changes the network density.

\subsection{Why Does Infrastructure Become Dynamic?}

The  mobile network infrastructure will become  stochastic  and the location of small cells cannot be pre-planned with the introduction of mobile cells.  Considering the   scenarios described above and shown in Fig.~\ref{scenario1}, we can list the major advantages of employing mobile or nomadic cells as follows.
\begin{itemize}
\item Mobile cells may be rapidly deployed to mitigate coverage holes without introducing site-acquisition costs.
\item Drone cells may facilitate ubiquitous coverage in rural areas.
\item Mobility of drones cells can be inline with the mobility of the end users providing a better approach for group mobility lowering the mobility management costs.
\item Mobile cells together with edge/fog computing may bring processing power closer to end users. Due to the reduction in the distance between UEs and base stations in small cells, power consumption will be reduced and higher data rates can be obtained by achieving high SINR.
\item Broadcast data rates can be improved especially for the UEs located at cell edges.
\end{itemize}
To attain these benefits, mobile cells have a huge potential to be employed in future networks. However, the mobility of cells~\cite{Yaliniz2016} is not the only factor that makes mobile networks and their infrastructure dynamic. The density of base stations in future mobile networks will vary in time and space because of additional technological advancements or reasons such as the following.
\begin{itemize}
\item User-controlled base stations (e.g., femtocells bought and controlled by end users): When  base stations are deployed in customer premises (such as homes), users may turn on or off them depending on  consumption requirements~\cite{6566960}.  
\item Green operation (e.g., sleep scheduling of base stations):  Base stations may employ duty-cycling for energy conservation. Depending on the employed duty-cycling scheme,  effective density of BSs will be different over  periods of time.
\item Incremental deployment: Gradual deployment of base stations will change the network density throughout the deployment time.
\item Loss of control and failures in the topology: Pre-planning and deterministic deployment is not probable any more. The operator may have to strictly comply with the constraints imposed by the urban structure; consequently, the deployment can be considered to be stochastic~\cite{Zahir2013}. 
\item Support for various verticals (e.g., automotive, health), multi-tenancy and various scenarios (e.g., megacities versus low-ARPU regions or sporadic events such as Olympics).
\end{itemize}

\begin{table*}[h]
\centering
\caption{Approaches for estimating  density of nodes in a network. Although some of the approaches are designed for ad hoc networks, they can generally be employed in any wireless network with minor modifications.}
\label{table:estimators}
\begin{tabular}{ | m{0.14\textwidth} | m{0.15\textwidth} | m{0.17\textwidth} | m{0.18\textwidth} | m{0.23\textwidth} | }
\hline
\textbf{Category}  & \textbf{Requirement} & \textbf{Advantages} & \textbf{Disadvantages} & \textbf{Related work} \\ \hline
\textbf{Location-based}            & The coordinates of  devices, location pre-awareness (e.g., GPS) & Ease of integration                                                                              & Extra energy consumption, errors in GPS measurement  & Node census~\cite{Raza2008}, density-adaptive routing~\cite{4489487}, priority-based stateless geo-routing~\cite{1542858}\\ \hline
\textbf{Neighborhood-based} & Monitoring and analyzing traffic, beaconing and neighbor discovery  & Existing functions in a stack can be employed                                                                              & Not scalable, limited to transmission range, accuracy depends on traffic & Traffic analysis~\cite{Raza2008} uses  network traffic, NEST~\cite{Iyer2012} uses beaconing\\ \hline
\textbf{Population-based} & Population census, pre-knowledge of nodes' positions & Convenient for simulation based studies & Suitable for small areas & Node census~\cite{Raza2008} uses population census,~\cite{Achtzehn} is applicable to homegenous  and inhomogenous Poisson Models\\ \hline
\textbf{Power-based} & Received signal strength or SINR measurements & Ease of integration, no other auxiliary function, or monitoring traffic of network & Sensitive to channel characteristics that may not be uniform in a field & Collaborative estimator~\cite{Onur2012,7343531} \\ \hline
\end{tabular}%
\end{table*}

\subsection{Why Present Architectures Will Fail?}

It is not possible today for present mobile communication networks to address these paradigm changes  because of their shortages and limitations~\cite{liyanage2015software}: 
\begin{itemize}
\item Lack of functions for mobile base stations (antennae): mobility management of base stations has not been foreseen in  standardization yet.
\item Inflexible architecture, static and manual configurations: when the infrastructure is  dynamic, it is clear static configurations will waste resources. Manual configurations make the network inflexible to dynamics in the topology and are subject to severe human errors. Software networks cast light onto this problems.
\item Lack of common control functions and interfaces: real-time and holistic management is  almost impossible because of vendor lock-in and  vendor-dedicated hardware and software components requiring trained/expert administrators. Network virtualization may help solve this problem.
\item Limited backhauling capacity and shortage of fiber infrastructure in developing countries. To fulfill the requirements of the aforementioned paradigm changes by overcoming the above limitations, heterogeneous networks consisting also of mobile, nomadic or stationary small cells can be a feasible approach. 
\end{itemize}

All in all, we can simply claim that cellular networks  start resembling ad-hoc networks. A distributed and self-* networking architecture is necessary.  Network density, whose estimators are presented in Section~\ref{sect:densityestimators},  is a crucial parameter since it substantially influences the network performance as we present in Section~\ref{sect:interman1}. We present in Section~\ref{challenges}  the challenges of adapting the network protocols to unpredictable density variations at run time to enhance the performance of the network  and the quality of service provided. The scarce resources such as energy and capacity will be wasted if the network protocols are configured without considering density.  Density-aware interference management is discussed as an application scenario of adaptation in Section \ref{sect:interman}.   We present an extensive list of research challenges in Section~\ref{sect:challenges}.

\section{Network Density}
\label{sect:interman1}
In dynamic networks, network density will change incessantly. We present how density can be estimated in the sequel. We qualitatively analyze the impact of density on performance. Then, we consider the BSs' density as another optimization parameter and show its impacts on  interference management  in dynamic networks in this section.

\subsection{Network Density Estimators}
\label{sect:densityestimators}
Network density is highly correlated with the location of base stations,  the neighborhood structure, the quality of received signals from other base stations or user equipments and population data. We can roughly categorize the network density estimation approaches as it can be seen in Table~\ref{table:estimators}. Location-based estimators  employ auxiliary positioning systems such as GPS that consume extra energy~\cite{Raza2008}~\cite{4489487}~\cite{1542858}. Neighborhood-based estimators, which are not scalable and suffer from inaccurate results, infer density from a census on  packet traffic~\cite{Iyer2012}. Power-based estimators combines the merits of location- and neighborhood-based estimators~\cite{Onur2012} although  received signal strength (RSS) is not a robust distance estimator. In cellular networks, spatial distribution of base stations (BSs) is vital for the analysis of connectivity, coverage  and  performance \cite{Achtzehn}. The proper adjustment of  spatial distribution and configuration of cells in simulators produces credible models important for capacity planning.  In \cite{Achtzehn}, the information of BS location obtained from different operators in Germany is used for finding out the utility and restrictions of population data as a base for the similar cellular deployments and it is shown that the density of  network is highly correlated to population data. They also figure out that relatively populated areas can be considered as a reasonable co-variate to model large-scale deployments. This study validates that predicting the number of BSs per unit area depending on the population density is sensible.  Proposing accurate density estimators is an open research challenge with a huge potential in stochastic geometry, especially for non-uniform deployments.

\subsection{Impact Analysis of Base Station Density}
\label{sect:impact}

A qualitative analysis of the impact of  base station density  on various mobile network parameters and performance measures is shown in Table~\ref{table:comparison}. The analysis is based on the following simple scenario. Assume a set of homogeneous base stations  are incrementally and randomly deployed in a field-of-interest. Suppose base stations are initially deployed sparsely, and service can only be given in a cluster of isolated coverage areas. As the density of base stations ($\lambda$) gradually increases (e.g., more and more base stations are deployed), isolated clusters merge and produce a huge cluster at a critical density ($\lambda_c$). At this stage, the global topology (macroscopic properties) of the network changes, and this phenomenon is called phase transition. 

\begin{table*}[]
\centering
\caption{The qualitative comparison of the impact of the density regime on  network  performance.}
\label{table:comparison}
\begin{tabular}{ | p{0.35\textwidth} | p{0.17\textwidth} | p{0.17\textwidth} | p{0.17\textwidth} | }\hline
                                                      	&	 \textbf{Sparse ($\lambda$ \textless $\lambda_c$)}                                      	&	 \textbf{Phase Transition} 	&	 \textbf{Dense ($\lambda$ \textgreater $\lambda_c$)}                     	\\ \hline
\textbf{Network capacity}                              	&	  low	&	 maximum                   	&	 low if no ICIC	\\ \hline
\textbf{Inter-cell Interference}                       	&	 low                                                                                    	&	 to be managed             	&	 high                                                                    	\\ \hline
\textbf{End-to-end throughput}                         	&	 low                                                                                    	&	 high                      	&	 low                                                                     	\\ \hline
\textbf{Coverage}                                      	&	 patchy                                                                                 	&	 resource-efficient        	&	 redundant                                                               	\\ \hline
\textbf{Mobility management}                           	&	 disruptive                                                                             	&	 optimal                   	&	 high cost                                                               	\\ \hline
\textbf{Number of relay base stations}                 	&	 few                                                                                    	&	 minimal                   	&	 large                                                                   	\\ \hline
\textbf{Possibility of multi-path routing}             	&	 none                                                                                   	&	 very low                  	&	 high                                                                    	\\ \hline
\textbf{Ratio of delay to sender-to-receiver distance} 	&	 scales sub-linearly                                                                    	&	                           	&	 scales linearly                                                         	\\ \hline
\textbf{Redundancy assisted topology control}          	&	 N/A                                                                                    	&	 possible                  	&	 possible                                                                	\\ \hline
\textbf{Resilience to failures}                        	&	 N/A                                                                                    	&	 low                       	&	 high                                                                    	\\ \hline
\end{tabular}
\end{table*}

The macro-behavior of the system below and above the critical density $\lambda_c$ is considerably different. There exists a giant component (coverage area) in the network consisting of active base stations in the dense networking regime where $\lambda > \lambda_c$. Whereas, the network is partitioned and there exists coverage holes in the sparse networking regime where $\lambda < \lambda_c$. The macroscopic behavior of the network changes from disrupted networking (i.e., isolated coverage areas having large capacity)  to degraded performance (full coverage with high interference) as the density increases. In this transition, at some density slightly larger than $\lambda_c$, resource-efficient (e.g., energy) operation of the network is possible. Therefore, the performance of the network is largely dependent on its topology that can be represented as a graph.

In graphs, phase transition is the concept where the probability of  the presence of a feature in a graph jumps from zero to one rapidly at a threshold value of the controllable parameter. The left- and right-hand sides of the threshold can be considered as static and chaotic regions. The region around the threshold is referred to as the phase transition region where innovations occur in a resource-efficient fashion. 

Take  transmit power adaptation as the example. The transmit power of base stations in a mobile network is a controllable parameter that may be employed to change the coverage area of the network. At a critical threshold of the transmit power, the connectivity of the network jumps from disconnected to highly-connected state. A level of transmit power less than the threshold causes a disconnected network, and the network is dysfunctional. Whereas, increasing the transmit power beyond the threshold causes a fully-connected network while increasing the interference and wasting  resources. Operating at the critical threshold facilitates resource-efficient networking. 

Similar phase transitions can be observed in many network design problems that are NP-hard such as drone cell placement~\cite{Lyu2016}. The complexity of such problems in the phase transition region surges. The centralized solutions of such problems do not scale in large networks. The network has to configure itself locally for resource efficiency.

The macro-behavior of the system below and above the critical density $\lambda_c$ is considerably different as shown in Table~\ref{table:comparison}. As the density of small cells increases, the coverage and capacity will grow due to a high level of spatial multiplexing. However, the capacity will eventually converge due to inter-cell interference in  dense networks when inter-cell interference control mechanisms (ICIC) are not employed. The global network capacity will be low in sparse networks due to the coverage holes and partitioning while the intra-cell capacity will be large due to the small amount of interference. The cost of mobility management escalates in dense networks due to the huge number of handovers. Whereas in sparse networks, mobility management would be disruptive due to the patchy coverage in the network. Multi-path routing and utilization of relay base stations also become infeasible in sparse networks. Topology control by exploiting the redundancy in the network is possible in dense networks. For instance, sleep scheduling of base stations can be employed considering the load in the network. The same fact also increases the resilience of the network to failures in  dense networks.

In dynamic  dense networks, collisions over random access channels, high congestion levels and inconstant capacities may be the significant challenges; whereas in sparse networks, partitioning is the key challenge. Dynamic networks have to collaborate locally for coverage preservation, mobility management, interference control and efficient resource allocation. However, the state-of-the-art architectures do not rely on localized cooperation. For carrying out those tasks in a density-adaptive fashion, base stations have to discover their neighborhood or estimate the density in an incessantly changing topology.  Edge computing can be a valuable technology towards this aim. 


As the cells become sporadic and their sizes changes, mobility management will be more cumbersome. When large cells are employed, paging costs are lower since the destination terminal is searched in fewer cells. When the cell sizes become small, paging consumes valuable in-band resources since a large number of cells are paged considering a constant location area mapping. Therefore, real-time decentralized management of cell sizes and coverage may have an adverse impact on the mobility management. 

\subsection{Density-aware Interference Management}
\label{sect:interman}

In  future networks, high-speed and ubiquitous connectivity will be an important demand that can be satisfied by  densification~\cite{pauli2010heterogeneous}~\cite{s17092077}. Network densification provide higher capacity with performing spatial reuse and less congestion with offloading~\cite{s17092077}. However, interference will be a significant problem to be tackled~\cite{pauli2010heterogeneous}~\cite{7010522}~\cite{s17092077}. Density-aware interference management will increase link capacity and spectral efficiency in  dynamic networks~\cite{s17092077}~\cite{6736751}.

\begin{figure}[h]
\centering
	\includegraphics [width=\columnwidth] {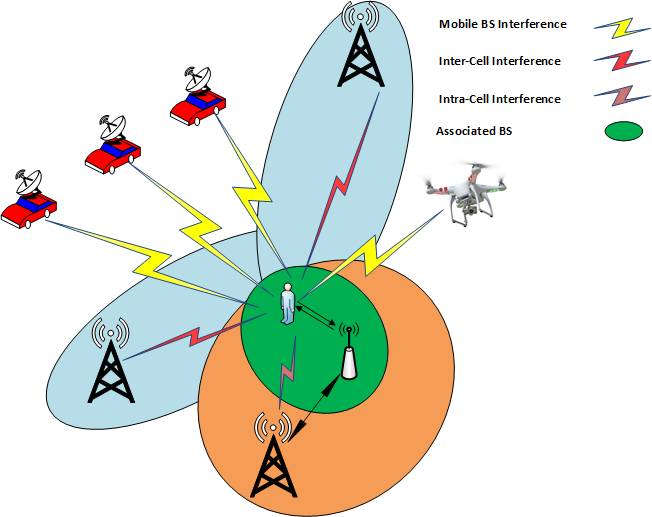}  
\caption{There will be various interferece sources in  dynamic networks.  What will be of notable importance is the interference from cells  on wings or wheels in future networks. }
\label{interfernce}
\end{figure}

In 4G mobile networks, if a UE  is located at cell edges, it can receive signals from multiple contiguous cells.  Inter-cell interference may originate from macro  or small base stations. In addition to inter-cell interference, different UEs  can interfere with each other as shown in Fig.~\ref{interfernce}. What will be of notable importance is the interference from nomadic or mobile cells in future networks. In Fig.~\ref{interfernce}, we present a scenario where a cell on wheels (mobile BS) interferes with a UE. This type of interference is the most difficult type if centralized solutions are to be employed.

Mobile operators may control  interference at three levels: at the Radio Access Network (RAN), between RAN and UE, and within UE.   
Various interference management approaches use both inter- and intra-cell communication methods~\cite{4907410}. Interference management methods include dynamic and static power control, resource partitioning techniques in time or frequency domain based on resource blanking in some cells by improving the signal quality of these resources in the neighbor cells, dynamic and static fractional frequency reuse, use of multiple receive and transmit antennae from different  sites, as well as antenna techniques based on space division multiple access in addition to beam-forming, multiple-input multiple-output (MIMO), some intelligent  and opportunistic scheduling techniques, considering network load and contention-based methods, and  interference cancellation techniques ~\cite{7010522}~\cite{6736751}~\cite{4907410}. Above all, network densification should be considered as an optimization parameter with these solutions. For instance, \cite{s17092077} uses network density as an optimization parameter  in addition to traffic load for interference management.

The concept of cloud radio access network (CRAN) provides centralization by accumulating base-band units in a virtualized resource pool. Software defined networking (SDN) and network function virtualization~\cite{6897914}~\cite{6882182} are the enablers. CRAN supports low latency and power consumption while facilitating coordinated multipoint (CoMP)  transmission and carrier aggregation that are highly correlated to interference management.

\section{Opportunities in Dynamic Networks}
\label{challenges}

Spectrum is a scarce resource in mobile communications. By densifying  mobile networks through employing small cells, higher signal-to-noise-plus-interference-ratios (SINR) can be achieved when interference cancellation and avoidance techniques are properly employed \cite{7840075}. However, transmitting signals with the objective of achieving larger bandwidth utilization and elevated data rates requires disproportionately high SINR for a given rate. Increasing the modulation level decreases the robustness of the system to noise and interference. Therefore, it is only possible to increase the modulation level in scenarios where high SINR can be obtained. Nomadic small cells that operate in the bandwidth-limited regime with low traffic levels are adequate candidates for increasing the SINR and facilitating higher-order modulations. 

In coordinated multi-point operation, BSs have to synchronize with each other over the X2 interface to transmit the same information to edge terminals. In this case, inter-cell interference becomes a productive phenomenon which is regarded and processed by the terminals using techniques to combat multipath fading \cite{Yassin2017}. With this approach, the broadcast rates are increased more in small cells in comparison to macro cells.

Channel quality may vary in time and frequency, and it is possible to measure the channel quality in both domains. In small cells where user mobility may be low, one may assume a dynamic channel in frequency while the channel quality does not vary considerably in time. In this case, user multiplexing over different carriers is a smarter option compared to time-domain channel scheduling. Depending on the physical layer dynamics, the radio link control has to support segmentation and concatenation of the frames. This is a clear requirement for a cross-layer design. The scheduler also has to deal with the inter-cell interference management. 

\begin{figure}[th!p]
\centering
\begin{subfigure}[b]{.45\textwidth}
	\includegraphics [width=\columnwidth] {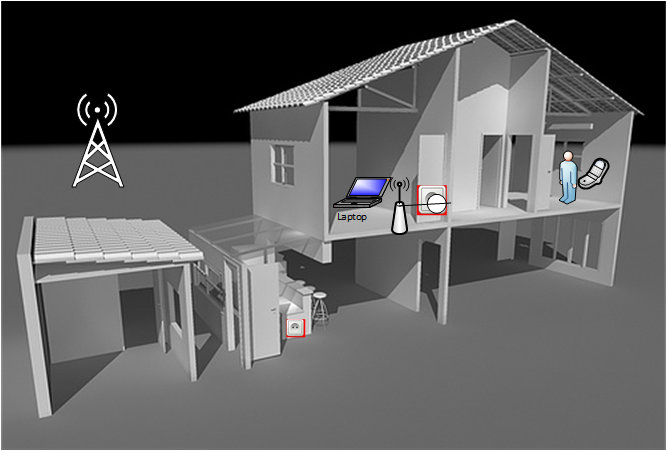} 
    \caption{BS is initially plugged to the mains power at the first floor.}
    \label{homea}
\end{subfigure}
\begin{subfigure}[b]{.45\textwidth}
	\includegraphics [width=\columnwidth] {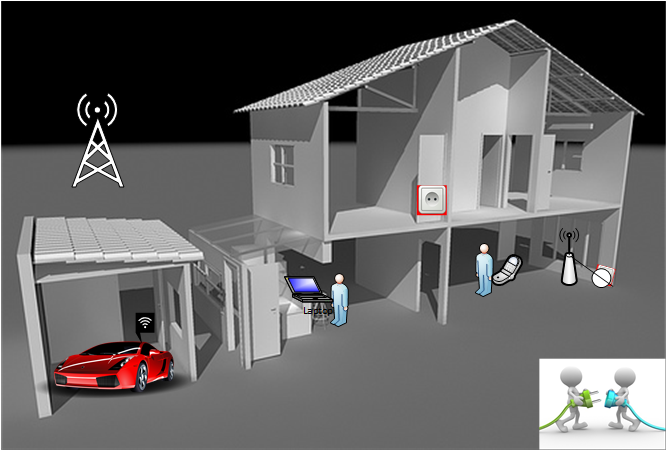} 
    \caption{The household moves the  BS to another part of the house.}
    \label{homeb}
\end{subfigure}
\caption{Mobile network operators may lose their control on the deployment and lifecycle of some small cells. Here, we illustrate a scenario where the household is able to change the location of a femto-cell deployed inside the house.}
\label{indoor}
\end{figure}

Wireless signals are considerably attenuated while penetrating inside the buildings in 3/4G mobile networks. The attenuation substantially decreases  signal-to-interference-plus-noise ratio (SINR)  and consequently data rates. Instead of outdoor deployments, indoor small cells may employ lower power levels and provide higher data rates compared to outdoor base stations. This scheme reduces energy consumption, improves quality of service (QoS) and experience (QoE), employs the spectrum efficiently, facilitate using licensed bands for home networking, lowers the level of electromagnetic radiations, minimizes the costs for the mobile operator and provides true ubiquity and coverage for subscribers \cite{LIN2017132}. However, operators lose their control on base station deployment.  As an example, there is an indoor femto BS  deployed in a house as shown in Fig.~\ref{homea}. Initially, the femto BS is operational at the first floor and users, instead of using outdoor BS, connect to the mobile network through the femto BS that can enhance  QoS and conserve energy. In Fig.~\ref{homeb},  the household decides to move the access point to the ground floor which  causes an uncontrolled BS failure for some time. Furthermore, this deployment change causes uncontrollable interference to neighboring houses after the BS becomes operational at its new location.

When multi-carrier modulation is employed, simultaneous transmission over sub-carriers may lead to greater deviations in instantaneous signal power and push amplifiers into the non-linear regions. This phenomenon leads to a larger amount of power consumption and dramatically increases the costs of amplifiers. The average transmit power can be lowered to avoid this problem. Small cells are adequate candidates toward this aim since the terminal-to-base distances in small cells are shorter. Mobile cells provides an advantage for getting the base stations closer and closer to users.  Then, large SINR facilitating higher-order modulations can be achieved.

Corruption due to frequency selectivity will lead to higher bit error rates and degrade the quality of the channel. Furthermore, there will be inter-symbol and inter-carrier interference. Frequency selectivity depends on the environment, especially obstructions and reflectors. Typically, less frequency selectivity is experienced in small  cells~\cite{dahlman20114g}. To combat frequency selectivity, cyclic-prefix insertion can be employed in multi-carrier modulation and the length of the prefix is a parameter to be adapted to the network density. The trade-off of cyclic prefix length is the robustness to time dispersion due to multipath fading and reduced data rates. Typically, time dispersion decreases as the cells become smaller and smaller. Therefore, a shorter prefix can be employed in small cells that increases data rates.

In macro cells, the downlink broadcast transmit powers should be controlled to be able to provide service to terminals close to the cell boundaries.  In dense networks,  broadcast data rates can be improved by reducing  distances to the terminals at  cell boundaries that increases the SINR of those terminals.  Doppler spread in multi-carrier transmission will destruct the orthogonality of the sub-carriers causing inter-carrier interference~\cite{dahlman20114g}. Frequency errors and phase noise also cause inter-carrier interference. Inter-carrier interference have a substantial impact in mobile cells compared to stationary cells. 

\begin{figure}[h!]
\centering
	\includegraphics [width=\columnwidth] {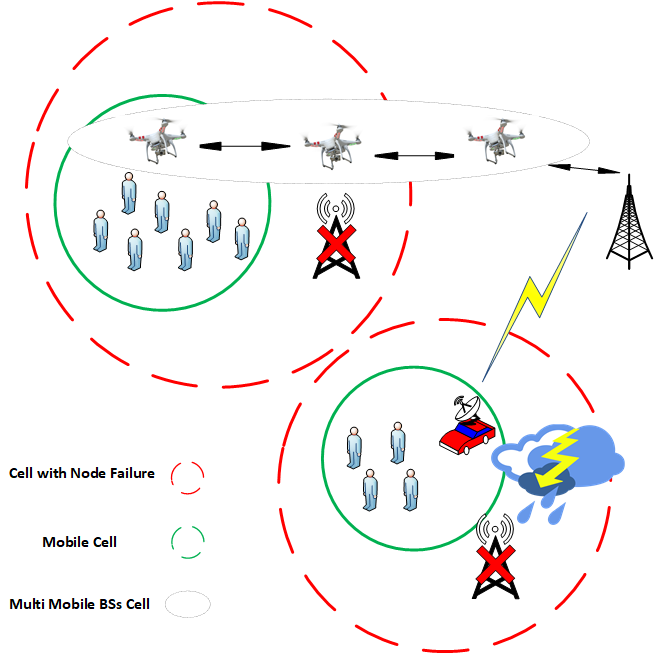}  
\caption{Ad hoc networking technologies can be employed to enlarge coverage of a mobile network. Mobile or stationary base stations may form an ad hoc infrastructure to backhaul traffic to the core network. Such dynamic infrastructures can also be established on demand; for example, when a stationary base station fails as we exemplify here. This feature may also play a critical role in public safety and disaster recovery.}
\label{multimobile}
\end{figure}

In stationary networks,   coverage  is restricted to the  range of BSs. However, by employing  mobile BSs that can collaborate with each other, the coverage  of a network can be substantially expanded. Beyond device-to-device communication, base stations can form an ad hoc network and establish a dynamic infrastructure to backhaul traffic to the core of the network as we show in Fig.~\ref{multimobile}. In case there are failures in backhaul or transport networks of a mobile network, mobile  or flying ad hoc networks (MANET or FANETS) of BSs can be employed to sustain communication. With this feature, we can enhance  communication reliability and maintain  connectivity through  mobile BSs \cite{Bekmezci2013}.

\section{Research Challenges of Dynamic Networks}
\label{sect:challenges}

The topology and coverage of the dynamic networks must be controlled since it significantly impacts the performance in terms of capacity, delay and resilience of the network to node and link failures. The topology depends on many controllable parameters and uncontrollable factors.  Interference, attenuation, environmental parameters such as obstructions, multipath propagation effects, fading and noise can be considered as uncontrollable factors which impact the link quality, and consequently the topology. These uncontrollable factors produce time- and space-variant links that are not predictable in advance. Cell mobility or presence may or may not be a controllable parameter that may sporadically cause blind spots or redundant coverage. Transmit power, antenna directivity, tilt or antennae count are the controllable parameters that can be used to change the network topology as required to make the network adaptive to density changes. Topology  and coverage control decisions should be given  autonomously based on the estimated density by the legitimate nodes or by mobile edge computing.

In one-cell frequency reuse, the same time-frequency resources can be reused in neighboring cells. Although this approach eases the network deployment and dramatically increases spectral efficiency, it may also cause large variations in SINR when proper inter-cell interference control (ICIC) mechanisms are not implemented~\cite{deb2014algorithms}.  Not only the neighboring small cells interfere with each other, but also the macro and small cells may interfere in heterogeneous networks. In dynamic networks where the density changes in time and space, density-aware dynamic interference management protocols are required.

On the boundary of large cells,  SINR will be anyhow low and a small amount of inter-carrier interference can still be neglected or tolerated. However, in small cells where high data rates are provided, the same amount of inter-carrier interference will have a larger negative impact. Therefore, density-aware mechanisms to mitigate the inter-carrier interference are also required.

Considering mobile base stations such as drone cells indicates the need for incorporation of delay-tolerant networking (DTN) concepts into mobile networking architectures. When backhauling is not possible, mobile base stations may have to manage the functions of the core network themselves. Lightweight evolving packet core (lightweight-EPC) and DTN may have to be considered for base stations on wheels or wings.

Mobility management is an integral part of mobile networks. However, managing the mobility of base stations is not considered. Not only users, but also the base stations have to be tracked and their locations have to be registered. Motion and deployment planning, self-control of mobility, handover management, virtual cells and new (dynamic) location area concepts are required and can be considered as open research challenges. Even when the users are stationary, handovers may be required when the base stations move. Furthermore, location area planning cannot be stationary any more since the infrastructure becomes dynamic.  

Accurate estimation of location plays a vital role in cooperative mobile BSs. In the current network, location estimation methods such as global positioning system mainly is used to calculate the coordinates of nomadic communication terminals and usually is sufficient to determine nodes’ locations. In case GPS is not available, by employing proximity based techniques or beacon nodes we can estimate the nodes’ coordination. Due to various mobility models of cells on wings or wheels, we need a highly accurate location estimator with a small  delay. GPS has 10 to 15 m error in location estimation and the location information can be received at one-second interval which may not applicable to multi mobile BSs and it can cause a collision among them under fast mobility or affect the channel conditions among them. To reduce  the estimation error,  assisted or differential GPS (AGPS or DGPS) can be used that can enhance the accuracy of estimation for about 10 cm by employing ground-based reference points \cite{5174293,AHN2009316}. To estimate  location faster, by equipping UAVs with an inertial measurement unit (IMU) which can be calibrated by help of GPS, location of mobile BSs can be retrieved faster and with higher  accuracy \cite{8073847}.

One of the most important parameters in any type of networks is latency which also needs to be considered in mobile BSs. Delay performance of mobile BSs  depends on the application. For instance, in  delay-sensitive applications such as reconnaissance, packets need to be delivered within a specific bound. When multiple mobile BSs (such as drone BSs) are deployed together to provide coverage, communication delay among those BSs need to be low to avoid any collisions among them. However, current protocols which are developed for MANETs may not applicable to  FANETs of BSs. Some applications of ad hoc networks of BSs may need  large bandwidth in addition to limited delay guarantees. The collaboration and communication of  mobile BSs with each other  require additional bandwidth. There are some important challenges  we need to consider for backhauling while allocating available bandwidth to mobile BSs. For instance, node failures and dynamic topology due to mobility of BSs will impact the achieved end-to-end capacity.

When mobile base stations are considered, channel models have to be revised. For cells on wings, air-to-ground,  ground-to-air and air-to-air channels have to be studied and modeled. Channel capacity will   be dynamic because of distance variation due to  mobility of BSs, reflection from ground (for drone cells),  variation of drone attitude, change of weather conditions in altitude, environmental clutter, interference from other BSs in three  dimensions (possibly four including time), and jamming by hostiles. All these additional constraints have to be considered in channel modeling of ad hoc networks of BSs.  When converged access/backhaul wireless links and the possibility of beyond-6 GHz operation are considered, channel modeling can be considered as an open research challenge. With the help of mobile base stations, there is a huge potential for relay base stations. In-band (converged access/backhaul) or out-of-band (radio links) relaying can be employed. The trade-offs between these approaches are to be evaluated.

Antenna design for dynamic networks is another challenge. While mobility management and neighbor discovery is easier when omnidirectional antennae are employed, energy  may be wasted. When directional antennae are employed, energy can be conserved or coverage can be expanded for a sector. For ad hoc networks of BSs, directional operation increases  spatial reuse in comparison to omnidirectional antennae. Furthermore, directional communication may decrease the number of hops thanks to antenna gains and reduce the end-to-end latency.  That is why, antenna design for dynamic networks is another important challenge.

In highly mobile dynamic ad hoc network of BSs, routing algorithm can affect the overall network performance. Employing proactive routing  requires constantly maintaining routing tables, and it may be costly under high mobility. On the other hand, determining routes before  packet transmission,  as in reactive routing, will be exhaustive. MANET routing protocols can be employed in dynamic networks.  Routing strategies, such as proactive, reactive or hybrid  approaches are extensively studied in the literature \cite{4224113, Karp:2000:GGP:345910.345953}. Position based routing protocols such as Greedy Perimeter Stateless Routing (GPSR) \cite{Karp:2000:GGP:345910.345953} may outperform in comparison to other  strategies.  However, in case of sparse deployment reliability can become a critical issue. If applications are sensitive to reliability, hybrid strategies can be employed \cite{5771206}.

Due to the tremendous pace of increasing multimedia services, caching may be required to decrease end-to-end delay and to efficiently use the scarce bandwidth. To achieve this goal, network operators employ  content caching at intermediate networked elements. One of the main issues of caching in  dynamic networks is determining the appropriate place for caching. By caching content close to network edges and users,  delay can be significantly reduced \cite{6736753}.  A challenge here  is to distribute caching to  network elements considering diversity, freshness, overhead of replications and their locations in a dynamic network.

Small cells may reduce CO$_2$-equivalent gas emissions per second. However, in ultra-dense networks the total sum may not be negligible. Furthermore, the new dimension to energy efficiency research will be trying to reduce the power consumed for mobility of base stations. Energy consumption,  CO$_2$-equivalent gas emissions and the impact of battery-driven operation of mobile base stations have to be carefully investigated.

Software-defined networking and network function virtualization are two distinct concept that may help implement dynamic networks. Centralized and stationary resource allocation may waste valuable resources. Distributing resource allocation tasks may enhance the efficiency. Through mobile edge computing hybrid approaches may be developed. End-to-end slicing will significantly be  more complex than the present approaches since the to-be-solved optimization problems  morphs with a higher frequency \cite{8025644}. One should also not forget the scalability requirements.

In dense deployments, UEs may camp on multiple base stations at the same time. Reliable end-to-end  communication requirements can then be addressed by employing multi-homed transport protocols not only in the control plane but also in the user/data plane~\cite{6503956}~\cite{1652245}~\cite{Mitharwal2014}. Multi-homing in user plane eases the mobility management burden.  As demonstrated in ~\cite{6503956} the transport layer multi-homed protocol has a better solution in order to provide reliable handover and connectivity.    Cell discovery, security, access scenarios have to be tackled in dynamic networks when multi-homing is employed.


\section{Conclusion}

With the invent of mobile base stations such as drone cells, not only the user's devices but also the elements in the infrastructure of the network has also become mobile introducing many novel and not-addressed challenges. A flexible and density-adaptive mobile communications architecture is required. However, there is a big research gap between the state of the art and the ambition of achieving a self-organized, adaptive and flexible networking architecture. In this paper, we present this gap  by presenting the paradigm changes in mobile communications and the consequences thereof. The existing architectures have severe limitations and shortages to be able to address the presented paradigm changes. We stress in this paper that density-aware and -adaptive  networking is crucial in future networks by presenting a qualitative and quantitative analysis of the impact of density on network performance. In this scope, we present the opportunities of dynamic networks as well as the challenges thereof.

All in all, one size protocols that are statically configured will not fit all scenarios in dynamic networks.  Robust interference management, coverage control and self-organization techniques that take mobile cells into account have to be developed. Such approaches may increase the cost of control.  Backhauling from cells on wheels or wings to the infrastructure may increase the load on and the cost of transport networks.  Traffic from mobile cells may overload the whole system if not controlled.	 Topology control and resource allocation become  very important challenges that cannot be easily addressed with the present inflexible management planes.

\section*{Acknowledgment}
This work is partially supported by TUBITAK under the grant number 215E127.


\begin{thebibliography}{10}
\providecommand{\url}[1]{#1}
\csname url@samestyle\endcsname
\providecommand{\newblock}{\relax}
\providecommand{\bibinfo}[2]{#2}
\providecommand{\BIBentrySTDinterwordspacing}{\spaceskip=0pt\relax}
\providecommand{\BIBentryALTinterwordstretchfactor}{4}
\providecommand{\BIBentryALTinterwordspacing}{\spaceskip=\fontdimen2\font plus
\BIBentryALTinterwordstretchfactor\fontdimen3\font minus
  \fontdimen4\font\relax}
\providecommand{\BIBforeignlanguage}[2]{{%
\expandafter\ifx\csname l@#1\endcsname\relax
\typeout{** WARNING: IEEEtran.bst: No hyphenation pattern has been}%
\typeout{** loaded for the language `#1'. Using the pattern for}%
\typeout{** the default language instead.}%
\else
\language=\csname l@#1\endcsname
\fi
#2}}
\providecommand{\BIBdecl}{\relax}
\BIBdecl

\bibitem{Yaliniz2016}
I.~Bor-Yaliniz and H.~Yanikomeroglu, ``The new frontier in {RAN} heterogeneity:
  Multi-tier drone-cells,'' \emph{IEEE Communications Magazine}, vol.~54,
  no.~11, pp. 48--55, November 2016.

\bibitem{6566960}
L.~Xie, Y.~Shi, Y.~T. Hou, W.~Lou, H.~D. Sherali, and S.~F. Midkiff, ``Bundling
  mobile base station and wireless energy transfer: Modeling and
  optimization,'' in \emph{Proc. of the IEEE INFOCOM}, April 2013, pp.
  1636--1644.

\bibitem{Zahir2013}
T.~Zahir, K.~Arshad, A.~Nakata, and K.~Moessner, ``Interference management in
  femtocells,'' \emph{IEEE Communications Surveys Tutorials}, vol.~15, no.~1,
  pp. 293--311, 2013.

\bibitem{Raza2008}
M.~H. Raza, L.~Hughes, and I.~Raza, ``Density: A context parameter of ad hoc
  networks,'' in \emph{Trends in Intelligent Systems and Computer Engineering},
  ser. Lecture Notes in Electrical Engineering, O.~Castillo, L.~Xu, and S.-I.
  Ao, Eds.\hskip 1em plus 0.5em minus 0.4em\relax Springer US, 2008, vol.~6,
  pp. 525--540.

\bibitem{4489487}
Z.~Li, Y.~Zhao, Y.~Cui, and D.~Xiang, ``A density adaptive routing protocol for
  large-scale ad hoc networks,'' in \emph{2008 IEEE Wireless Communications and
  Networking Conference}, March 2008, pp. 2597--2602.

\bibitem{1542858}
Y.~Xu, W.-C. Lee, J.~Xu, and G.~Mitchell, ``{PSGR}: priority-based stateless
  geo-routing in wireless sensor networks,'' in \emph{Proc. of the IEEE MASS},
  Nov. 2005, pp. 680--688.

\bibitem{Iyer2012}
V.~Iyer, A.~Loukas, and S.~Dulman, ``Nest : A practical algorithm for
  neighborhood discovery in dynamic wireless networks using adaptive
  beaconing,'' Delft University of Technology, Tech. Rep., 2012.

\bibitem{Achtzehn}
A.~Achtzehn, J.~Riihijarvi, and P.~Mahonen, ``Large-scale cellular network
  modeling from population data: An empirical analysis,'' \emph{IEEE
  Communications Letters}, vol.~20, no.~11, pp. 2292--2295, Nov 2016.

\bibitem{Onur2012}
E.~Onur, Y.~Durmus, and I.~Niemegeers, ``Cooperative density estimation in
  random wireless ad hoc networks,'' \emph{IEEE Communications Letters},
  vol.~16, no.~3, pp. 331--333, Mar 2012.

\bibitem{7343531}
A.~Eroğlu, E.~Onur, and H.~Oğuztüzün, ``Estimating density of wireless
  networks in practice,'' in \emph{2015 IEEE 26th Annual International
  Symposium on Personal, Indoor, and Mobile Radio Communications ({PIMRC})},
  Aug 2015, pp. 1476--1480.

\bibitem{liyanage2015software}
M.~Liyanage, A.~Gurtov, and M.~Ylianttila, \emph{Software Defined Mobile
  Networks (SDMN): Beyond {LTE} Network Architecture}.\hskip 1em plus 0.5em
  minus 0.4em\relax John Wiley \& Sons, 2015.

\bibitem{Lyu2016}
J.~Lyu, Y.~Zeng, R.~Zhang, and T.~J. Lim, ``Placement optimization of
  {UAV}-mounted mobile base stations,'' \emph{IEEE Communications Letters},
  vol.~PP, no.~99, pp. 1--1, 2016.

\bibitem{pauli2010heterogeneous}
V.~Pauli, J.~D. Naranjo, and E.~Seidel, ``Heterogeneous {LTE} networks and
  inter-cell interference coordination,'' \emph{Nomor Research GmBH}, pp. 1--9,
  2010.

\bibitem{s17092077}
\BIBentryALTinterwordspacing
J.~Feng and Z.~Feng, ``Optimal base station density of dense network: From the
  viewpoint of interference and load,'' \emph{Sensors}, vol.~17, no.~9, pp.
  2077--2095, Sep 2017. [Online]. Available:
  \url{http://www.mdpi.com/1424-8220/17/9/2077}
\BIBentrySTDinterwordspacing

\bibitem{7010522}
B.~Soret, K.~I. Pedersen, N.~T.~K. Jørgensen, and V.~Fernández-López,
  ``Interference coordination for dense wireless networks,'' \emph{IEEE
  Communications Magazine}, vol.~53, no.~1, pp. 102--109, January 2015.

\bibitem{6736751}
S.~Hong, J.~Brand, J.~I. Choi, M.~Jain, J.~Mehlman, S.~Katti, and P.~Levis,
  ``Applications of self-interference cancellation in {5G} and beyond,''
  \emph{IEEE Communications Magazine}, vol.~52, no.~2, pp. 114--121, February
  2014.

\bibitem{4907410}
G.~Boudreau, J.~Panicker, N.~Guo, R.~Chang, N.~Wang, and S.~Vrzic,
  ``Interference coordination and cancellation for 4g networks,'' \emph{IEEE
  Communications Magazine}, vol.~47, no.~4, pp. 74--81, April 2009.

\bibitem{6897914}
A.~Checko, H.~L. Christiansen, Y.~Yan, L.~Scolari, G.~Kardaras, M.~S. Berger,
  and L.~Dittmann, ``Cloud {RAN} for mobile networks; a technology overview,''
  \emph{IEEE Communications Surveys Tutorials}, vol.~17, no.~1, pp. 405--426,
  Firstquarter 2015.

\bibitem{6882182}
C.~L. I, J.~Huang, R.~Duan, C.~Cui, J.~. Jiang, and L.~Li, ``Recent progress on
  {C}-{RAN} centralization and cloudification,'' \emph{IEEE Access}, vol.~2,
  pp. 1030--1039, 2014.

\bibitem{7840075}
K.~Song, B.~Ji, Y.~Huang, M.~Xiao, and L.~Yang, ``Performance analysis of
  heterogeneous networks with interference cancellation,'' \emph{IEEE
  Transactions on Vehicular Technology}, vol.~66, no.~8, pp. 6969--6981, Aug
  2017.

\bibitem{Yassin2017}
\BIBentryALTinterwordspacing
M.~Yassin, M.~A. AboulHassan, S.~Lahoud, M.~Ibrahim, D.~Mezher, B.~Cousin, and
  E.~A. Sourour, ``Survey of {ICIC} techniques in {LTE} networks under various
  mobile environment parameters,'' \emph{Wireless Networks}, vol.~23, no.~2,
  pp. 403--418, Feb 2017. [Online]. Available:
  \url{https://doi.org/10.1007/s11276-015-1165-z}
\BIBentrySTDinterwordspacing

\bibitem{LIN2017132}
\BIBentryALTinterwordspacing
K.-H. Lin, C.-H. Tsai, J.-W. Chang, Y.-C. Chen, H.-Y. Wei, and F.-M. Yeh,
  ``Max-throughput interference avoidance mechanism for indoor self-organizing
  small cell networks,'' \emph{ICT Express}, vol.~3, no.~3, pp. 132 -- 136,
  2017. [Online]. Available:
  \url{http://www.sciencedirect.com/science/article/pii/S2405959516301813}
\BIBentrySTDinterwordspacing

\bibitem{dahlman20114g}
E.~Dahlman, S.~Parkvall, and J.~Skold, \emph{{4G}: {LTE}/{LTE-Advanced} for
  Mobile Broadband}, 1st~ed.\hskip 1em plus 0.5em minus 0.4em\relax Academic
  Press, 2011.

\bibitem{Bekmezci2013}
I.~Bekmezci, O.~K. Sahingoz, and S.~Temel, ``{Flying {Ad-Hoc} Networks
  ({FANET}s): A survey},'' \emph{Ad Hoc Networks}, vol.~11, no.~3, pp.
  1254--1270, Jan 2013.

\bibitem{deb2014algorithms}
S.~Deb, P.~Monogioudis, J.~Miernik, and J.~P. Seymour, ``Algorithms for
  enhanced inter-cell interference coordination ({eICIC}) in {LTE} {HetNets},''
  \emph{IEEE/ACM Transactions on Networking}, vol.~22, no.~1, pp. 137--150, Feb
  2014.

\bibitem{5174293}
A.~K. s.~Wong, T.~K. Woo, A.~T.~L. Lee, X.~Xiao, V.~W.~H. Luk, and K.~W. Cheng,
  ``An {AGPS}-based elderly tracking system,'' in \emph{2009 First
  International Conference on Ubiquitous and Future Networks}, June 2009, pp.
  100--105.

\bibitem{AHN2009316}
\BIBentryALTinterwordspacing
H.-S. Ahn and C.-H. Won, ``{DGPS/IMU} integration-based geolocation system:
  Airborne experimental test results,'' \emph{Aerospace Science and
  Technology}, vol.~13, no.~6, pp. 316 -- 324, 2009. [Online]. Available:
  \url{http://www.sciencedirect.com/science/article/pii/S1270963809000261}
\BIBentrySTDinterwordspacing

\bibitem{8073847}
J.~Huang, Z.~Huang, and K.~Chen, ``Combining low-cost inertial measurement unit
  ({IMU}) and deep learning algorithm for predicting vehicle attitude,'' in
  \emph{2017 IEEE Conference on Dependable and Secure Computing}, Aug 2017, pp.
  237--239.

\bibitem{4224113}
M.~T. Hyland, B.~E. Mullins, R.~O. Baldwin, and M.~A. Temple,
  ``Simulation-based performance evaluation of mobile {Ad Hoc} routing
  protocols in a swarm of unmanned aerial vehicles,'' in \emph{Advanced
  Information Networking and Applications Workshops, 2007, AINAW '07. 21st
  International Conference on}, vol.~2, May 2007, pp. 249--256.

\bibitem{Karp:2000:GGP:345910.345953}
\BIBentryALTinterwordspacing
B.~Karp and H.~T. Kung, ``{GPSR}: Greedy perimeter stateless routing for
  wireless networks,'' in \emph{Proceedings of the 6th Annual International
  Conference on Mobile Computing and Networking}, ser. MobiCom '00.\hskip 1em
  plus 0.5em minus 0.4em\relax New York, NY, USA: ACM, 2000, pp. 243--254.
  [Online]. Available: \url{http://doi.acm.org/10.1145/345910.345953}
\BIBentrySTDinterwordspacing

\bibitem{5771206}
R.~Shirani, M.~St-Hilaire, T.~Kunz, Y.~Zhou, J.~Li, and L.~Lamont, ``The
  performance of greedy geographic forwarding in unmanned aeronautical {Ad-Hoc}
  networks,'' in \emph{2011 Ninth Annual Communication Networks and Services
  Research Conference}, May 2011, pp. 161--166.

\bibitem{6736753}
X.~Wang, M.~Chen, T.~Taleb, A.~Ksentini, and V.~C.~M. Leung, ``Cache in the
  air: exploiting content caching and delivery techniques for {5G} systems,''
  \emph{IEEE Communications Magazine}, vol.~52, no.~2, pp. 131--139, February
  2014.

\bibitem{8025644}
E.~J. Kitindi, S.~Fu, Y.~Jia, A.~Kabir, and Y.~Wang, ``Wireless network
  virtualization with {SDN} and {C-RAN} for {5G} networks: Requirements,
  opportunities, and challenges,'' \emph{IEEE Access}, vol.~5, pp.
  19\,099--19\,115, 2017.

\bibitem{6503956}
M.~S. Hossain, M.~Atiquzzaman, and W.~Ivancic, ``Performance comparison between
  multihomed network mobility protocols,'' in \emph{2012 IEEE Global
  Communications Conference (GLOBECOM)}, Dec 2012, pp. 5260--5265.

\bibitem{1652245}
K.~Sethom, O.~Hamza, and H.~Afifi, ``Smart multi-homed mobile networks,'' in
  \emph{2006 ACS/IEEE International Conference on Pervasive Services}, June
  2006, pp. 291--294.

\bibitem{Mitharwal2014}
P.~Mitharwal, C.~Lohr, and A.~Gravey, \emph{Survey on Network Interface
  Selection in Multihomed Mobile Networks}.\hskip 1em plus 0.5em minus
  0.4em\relax Cham: Springer International Publishing, 2014, pp. 134--146.

\end{thebibliography}


\end{document}